# Packing Concave Molecules in Crystals and Amorphous Solids: On the Connection between Shape and Local Structure

Cerridwen Jennings, Malcolm Ramsay, Toby Hudson and Peter Harrowell

*School of Chemistry, University of Sydney, Sydney NSW 2006 Australia*

Abstract

The structure of the densest crystal packings is determined for a variety of concave shapes in 2D constructed by the overlap of two or three disks. The maximum contact number per particle pair is defined and proposed as a useful means of categorizing particle shape. We demonstrate that the densest packed crystal exhibits a maximum in the number of contacts per particle but does not necessarily include particle pairs with the maximum contact number. In contrast, amorphous structures, generated by energy minimization of high temperature liquids, typically do include maximum contact pairs. The amorphous structures exhibit a large number of contacts per particle corresponding to over-constrained structures. Possible consequences of this over-constraint are discussed.

## 1. Introduction

Since the 1960's, the statistics of packing hard objects in space has occupied a central place in the theoretical description of liquids and crystals [1]. Impenetrability is the generic consequence of the short range repulsions between particles and the guarantor of the stability of matter [2]. In this context, hard spheres have provided a generic reference state for atomic liquids. In 1968, Verlet [3] confirmed that the short range structure of a model of liquid argon was well reproduced by the analogous correlations in a liquid of hard spheres. The maximum density packing of hard spheres – either a single size [4] or a mixture of two sizes [5,6] – is crystalline, an important result that extends the utility of the hard sphere reference state to include the crystal phase and the freezing transition. For this reason, the dense packings – both crystalline and amorphous – of hard spheres and mixtures of hard spheres have been the subject of a considerable literature [4,5,6]. Non-spherical shapes have, likewise, received considerable attention as references for molecular liquids (including liquid crystals) and, more recently, anisotropic colloidal particles. The list of convex shapes whose dense packings have been studied includes ellipsoids [7], spherocylinders [8], cut spheres [9] and regular polyhedra [10,11].

There have been relatively few theoretical studies of the nature of dense packing of concave particles, despite the fact that any reasonable description of the short range repulsions between molecules must, typically, entail concavities. In 2011, de Graaf *et al* presented a preliminary study which included the dense packing of a sample of concave shapes [11]. Dimers consisting of touching spheres [12] sample a subset of binary spheres structures so that the concavity of the dimer becomes equivalent to the packing of the constituent spheres. Milinkovic *et al* [13] have reported that the equilibrium solid phases for dimers where the component spheres are allowed to overlap are similar to those found previously for the touching spheres. Atkinson *et al* [14] have reported the packing of a selection of concave shapes in 2D: crosses, concave triangles and crescents. Advances in the manipulation of colloid particles permits the synthesis of suspensions consisting of a single anisotropic particle. Among these new shaped colloids, there are a number of with concavities [15-18] and anisotropic attractions [19-21]. The potential novel packings arising from concave particles have been studied in some detail for the case of 'indented' colloids by Ashton et al [22]. In addition to the steric effects of the concavity, these workers have explored the added flexibility of arising from combining concavities with depletion interactions.

By way of a general organization principle, Torquato and Jiao [23] proposed that a packing of concave polyhedra with central symmetry consisting of the densest lattice packing (i.e. a single particle per unit cell) would provide a useful lower bound on the packing density. For concave particles without central symmetry, they propose that particles would pair so that the pair now possessed an inversion centre (i.e. a particle paired with another rotated through $180^o$) and, following from their previous postulate, packing this pair on a lattice (i.e. with 2 particles per unit cell) would also provide a useful lower bound for the crystal density . We shall consider the utility of these bounds in the light of our calculations below.

A particular challenge in any study of the condensed behaviour of concave shapes is to come up a classification of such shapes that is both sufficiently compact and physically relevant to be of use. One approach, applied in 2D [24] and 3D [11] is to consider two circles – a *circumscribing* circle defined as the smallest circle capable of completely enclosing the particle and an *inscribed* circle defined as the largest circle that lies entirely within the particle. The ratio of the difference in these two radii over the radius of the circumscribing circle provides a measure of shape that is zero for a circle and increases with anisotropy of the particle. Tracey, Widmer-Cooper and Hudson [25] have examined the correlation of this and a wide range of other possible measures in terms of their value as a predictor of the packing fraction of the densest crystal packing. In this paper we shall examine the utility of the classifying shape in terms of the maximum number of contacts that can be made between a pair of congruent particles.

In this study we shall restrict our attention to the packing of molecule-like concave shapes in 2D with these shapes generated by overlapping disks as shown in Fig. 1.

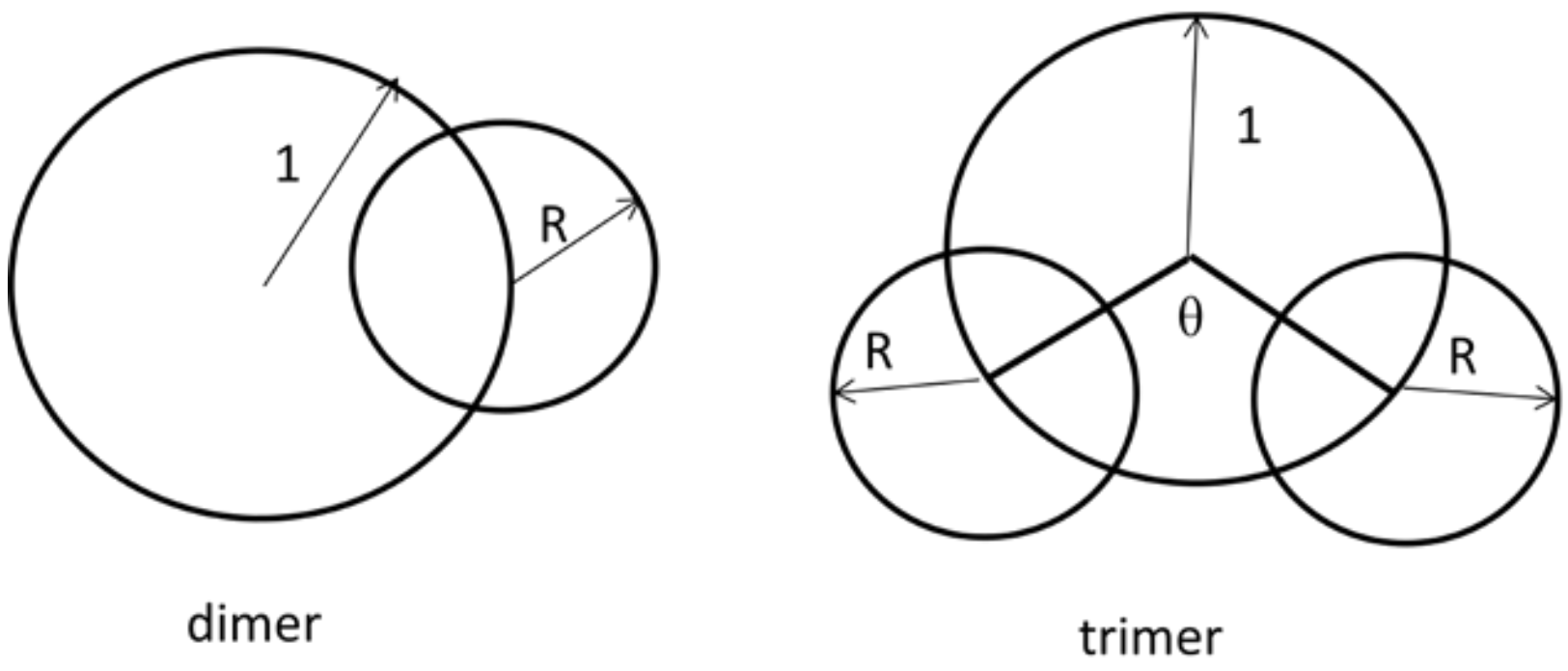

**Figure 1.** The construction of the dimer and trimer shapes and the identification of the shape variables: R for the dimer and (R,θ) for the trimer.

## 2. Particle Shape and the Maximum Contact Number per Particle Pair

A defining feature of a concave particle is that it is capable of making multiple contacts with a neighbouring particle. As contacts represent mechanical constraints, this aspect of concavity is likely to play an important role in determining the physical properties of condensed phases. The variability in the number of contacts possible between a pair of concave particles suggests that the *maximum contact number* (MCN) for a pair of particles might provide a useful and general means of categorizing shape. The MCN is a well defined property of a shape that we have determined by a systematic enumeration of contact numbers over all possible orientations of a pair of particles. Note that the MCN for a given shape might correspond to a unique pairing of the particles or it might be achievable by a number of different mutual arrangements of the two particles. The MCN is an integer $\geq 1$. For arbitrary shapes, the MCN will also be bounded above so that in 2D we expect MCN $\leq 4$ because once the contacts have reached 4 there is no further freedom to reorganise the particle pairs so that any additional contacts would only be possible if they were specifically designed into the shape for the particle. The categories of particle shape that the MCN resolved are illustrated in Figure 2.

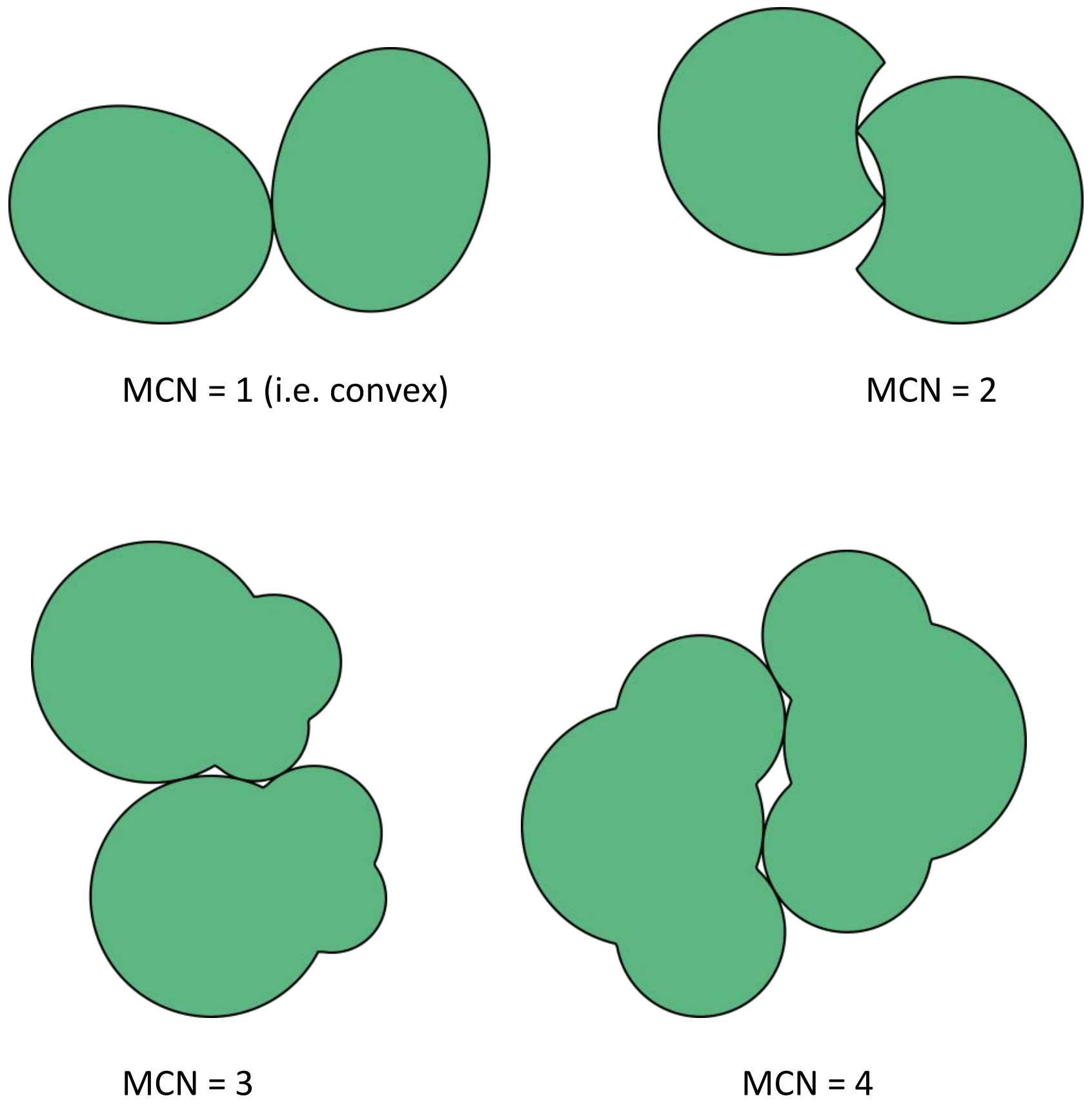

**Figure 2.** Examples of particle shapes in 2D associated with the different values of the maximum contact number MCN.

*We find that a rotated particle pair (i.e. a pair of particles in contact related by a 180$^o$ rotation) can always be found within the set of contact pairs exhibiting the maximum contact number.* Whether this observation is associated with the already noted preference for double lattice packings will depend on the prevalence of MCN pairs in dense packing, a question we shall address in the following Sections.

## 3. Densest Crystal Packing

The structural searches for the maximum density crystals, reported in this Section, have been carried out using the Isopointal Search Algorithm introduced by Hudson and Harrowell [26] and implemented for anisotropic particles by Elias and Hudson [27]. This search involves a systematic calculation of the maximum density packings of all possible 2D crystals that satisfy a given restriction on the particles in the unit cell. The possible sites in a unit cell can be categorized on the basis of the symmetry elements they are associated with e.g. a site that lies on a reflection plane can be distinguished from one that lies specifically on a centre of inversion or one that occupies a general position without any special symmetry. Each crystal space group is characterized by a list of the number of different types of site that can be distinguished in this manner. These sites are known as the Wyckoff positions for that space group [28]. In the implementation of the Isopointal Search Algorithm [26] used in this paper we restrict the search to the case where only a single Wyckoff position is occupied. Unit cells consisting of 2 or more particles are allowed under this search restriction as long as the particles are related to one other by the symmetry operations of the space group. Steed [29] has noted that ~ 92% of reported molecular crystal structures satisfy this condition. The result of this search is the maximum possible density for each of the 17 space groups.

The use of packing arguments to rationalise the structure of molecular crystals is widely associated with Kitaigorodsky [30], who argued that close packing in 2D required a coordination number of 6 and that this coordination number was only possible for molecules of arbitrary shape (i.e. without any symmetry) in 4 of the 17 space groups: p1, p2, pg and p2gg. (Examples of these space groups are provided in the Appendix.) This substantial reduction in the space of possible crystals is reflected in the experimental data set of molecular crystals [29,31].

The density of a packing is typically reported as a reduced quantity known as the packing fraction $\phi$, where $\phi$ = *volume occupied by particles/total volume.* Hard disks pack in a triangular lattice with a packing fraction $\phi = \frac{\pi}{2\sqrt{3}} \approx 0.907$ . In 1990, Kuperberg and Kuperberg [32] proved that the packing of convex bodies of arbitrary symmetry in 2D could always achieve densities in excess of $\phi = \sqrt{3}/2 \approx 0.866$ using ‘double lattice’ packings – i.e. packings consisting of adjacent layers related by a 180$^o$ rotation. While this lower bound is not that impressive (it is reasonable to suppose that the packing of hard disks provides a better lower bound for the packing of shapes constructed from overlapping disks), the identification of double lattice packings as a generally useful candidate for the optimal packing of concave shapes is supported by our results below. Double lattice packings can also be considered to be a lattice of rotated particle pairs – a pairing that results in a centre of inversion as discussed in ref. [15]. In the case of a chiral shape, the double lattice condition would impose a significant restriction on the space groups with only p2 being possible. For achiral shapes such as those considered here, more space groups are possible, at least in principle. The group p2 is still allowed, pg is allowed (but only if we relax the 180$^{o}$ rotation feature of the double lattice construction) and the groups p2mm and c2mm are allowed but with severely constrained packing efficiencies (4 and 5 neighbours, respectively). (Examples of these various structures are provided in the Appendix.)

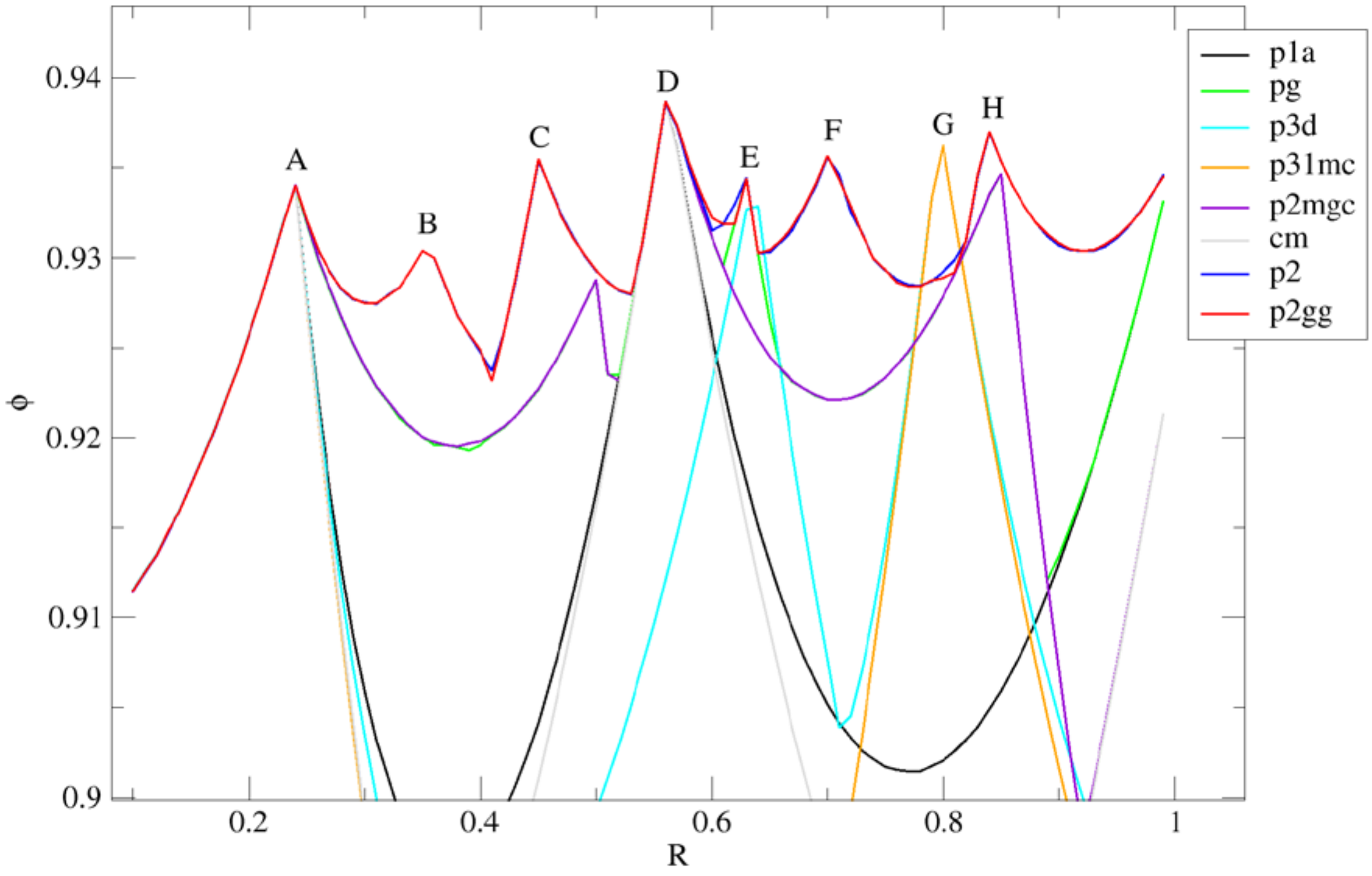

**Figure 3.** The packing fraction ϕ of dimer crystals as a function of R, the radius of the smaller component disk (see Fig. 1).

In Fig. 3 we present the maximum packing fraction for each crystal space group for dimers as a function of the minor radius R. (Consistently low density space groups have been omitted.) Note that the MCN = 3 for the dimers across the entire range of R. There are a number of points worth noting regarding the dense packing of dimers.

i) The packing fractions of the dimers are generally higher than the value for the densest packing of discs (i.e. $\phi = 0.907$). The origin of this increase is perhaps clearest when the optimum structures in the range $R \leq 0.25$. Over this range, the optimal packing is based on the triangular arrangement of the disks with the small protuberances directed so as to occupy the voids of that reference lattice, as shown in Fig. 4.

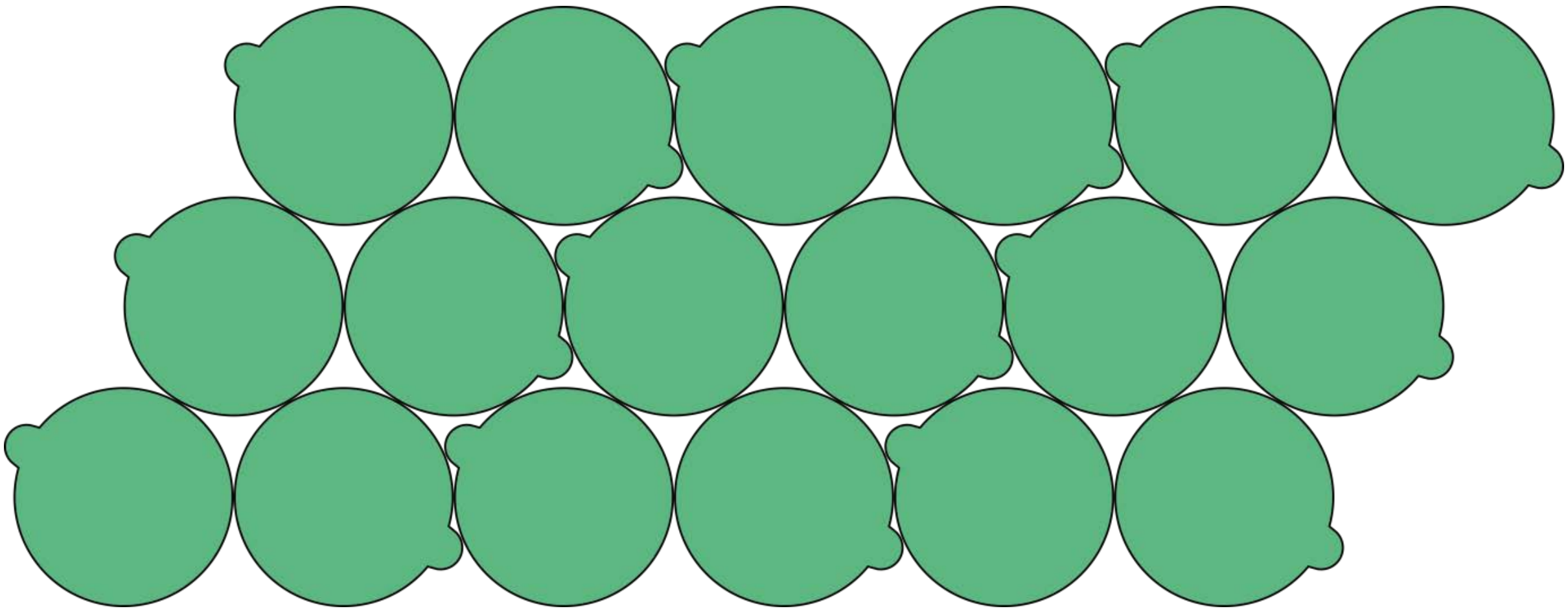

**Figure 4.** One of the degenerate densest packing of dimers with R=0.2. Note the triangular packing of the larger disks.

ii) The densest packings are overwhelmingly dominated by the space groups p2 and p2gg. We find almost perfect degeneracy of the optimized p2 and p2gg crystals across the entire range of R, the radius of the smaller disk. The origin of the frequent degeneracy of these two structures, which are often quite distinct, remains an open question.

iii) The peaks in the packing fraction correspond to shapes that achieve an optimal number of contacts per particle. In the case of the dimers, this number appears to be 10 contacts per particle. The structures corresponding to these peaks in packing fraction are depicted in Fig. 5 using the labels provided in Fig. 3.

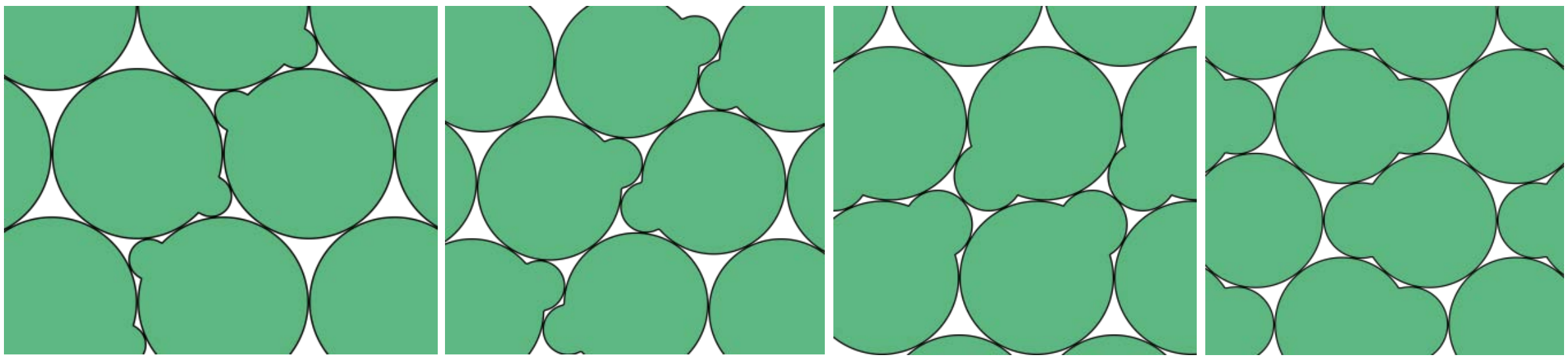

**A** R=0.24 ϕ=0.9340 **B** R=0.35 ϕ=0.9354 **C** R=0.45 p=0.9354 **D** R=0.56 ϕ=0.9386

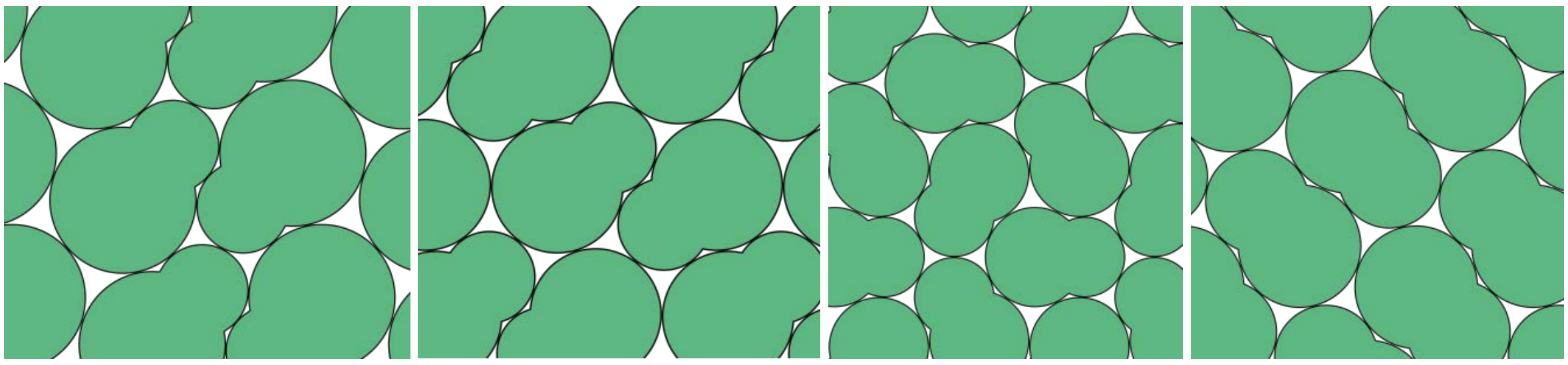

**E** R=0.63 ϕ=0.9344 **F** R=0.70 ϕ=0.9356 **G** R=0.80 ϕ=0.9362 **H** R=0.84 ϕ=0.9370

**Figure 5.** Densest packings of dimers at the peaks in the packing fraction as indicated in Fig. 3. All of these packings are characterised by 10 contacts per particle. Structure G has symmetry p31mc while the rest are p2 (with a p2gg structure equal in density). Note that only the 3-fold structure (i.e. G) fails to employ the MCN pairing of 3 contacts.

This coincidence between high density packings and a maximum number of contacts per particle is an important observation, especially if confirmed for some general range of shapes. An optimal number of contacts is generally expected to be a singular function of any continuously variable related to shape, i.e. any variation of a shape away from one of these packing peaks will force apart one or more of the contacts. This accounts for the cusp-like peaks in these packing fraction curves. The structures on either immediate side of a peak are distortions of the ideal structure while the minima in the packing fraction curves correspond to switches from one structure type to another.

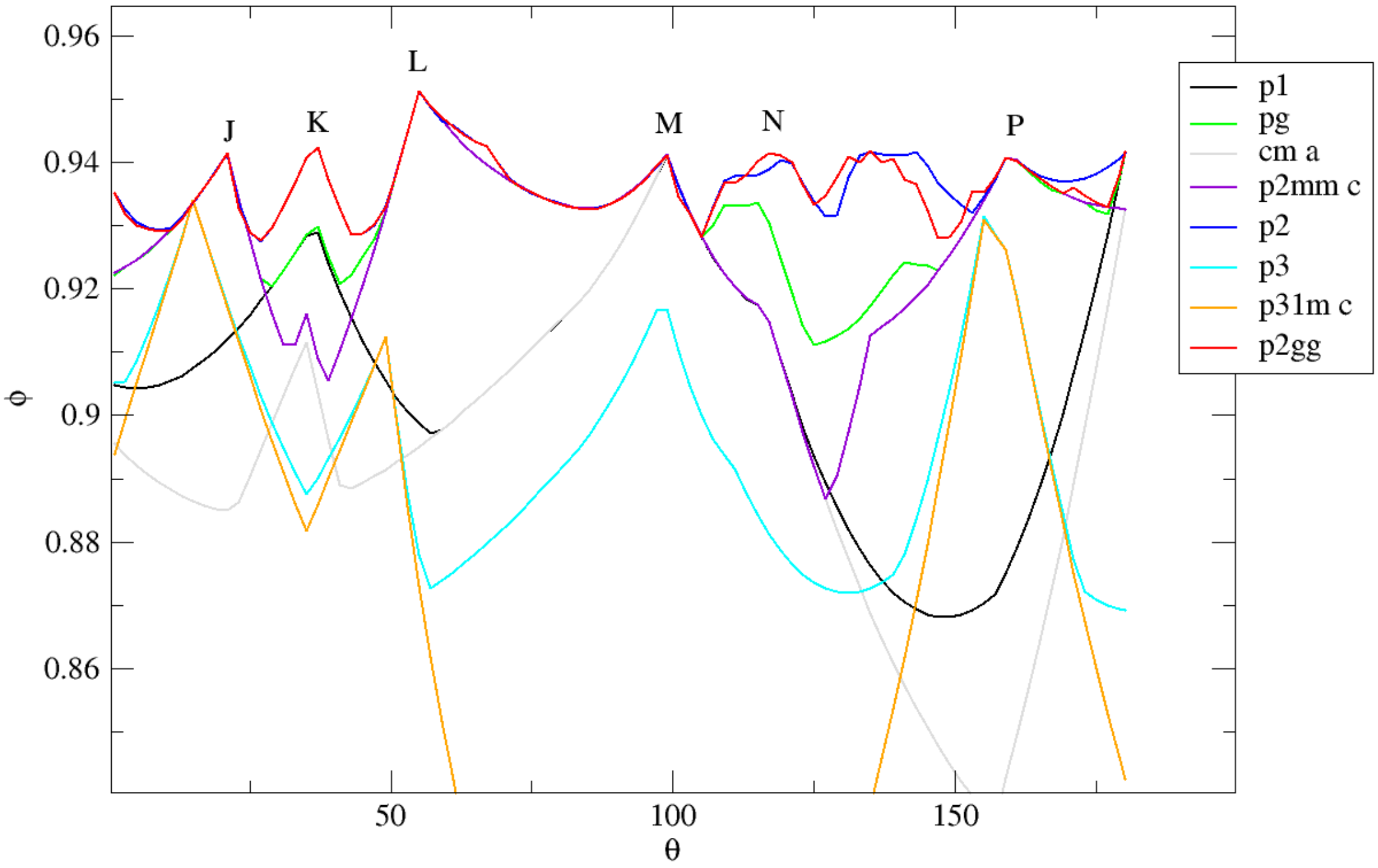

**Figure 6.** The packing fraction of the trimer with R= 0.7 as a function of the bond angle.

Turning to the trimer, we present the maximum packing fraction for shapes characterised by R=0.7 for a range of the bond angle θ in Fig. 6. Many of the features of these packing fraction curves have already been noted for the dimers. The maximum packing fractions, considerably larger than the disk value, are dominated by the p2 and p2gg structures. Again we find these structures to be near degenerate in density across the range of bond angles up to ~ $111^{o}$ despite the clear structural differences (see Fig. 7). For greater bond angles we find the packing fractions of the p2 and p2gg structures differ. We also find that the cusp-like peaks indicated in Fig. 6 whose structures are depicted in Fig. 8. For all values of θ, our densest packed structure is characterized by a maximum in the contact number per particle relative to the best packings of the other space groups. As shown in Fig. 9, the packings associated with peaks in the packing fraction are also associated with maxima in the contact number with respect to the variation of particle shape (i.e. a change in the bond angle) -12 contacts per particle in all cases except for the optimal structure at $\theta = 143^{0}$ where we find 13 contacts per particle. These local maxima in the contact number appear to be singular functions of the shape.

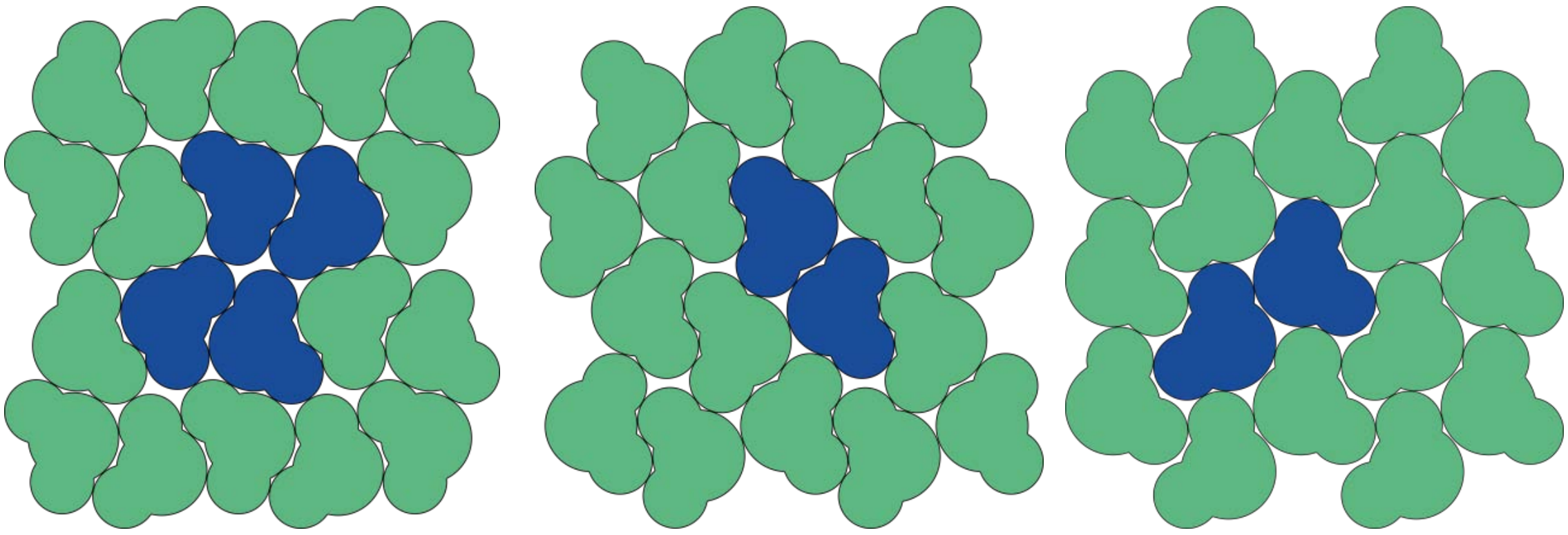

p2gg $\phi$=0.9366 p2 $\phi$ = 0.9376 pg $\phi$=0.9330

**Figure 7.** Comparison of the structures for the three densest crystals for the symmetric trimer with R=0.7 and $\theta = 111^o$. Note that the p2gg and p2 structures, while essentially indistinguishable in terms of density, have quite distinct structures.

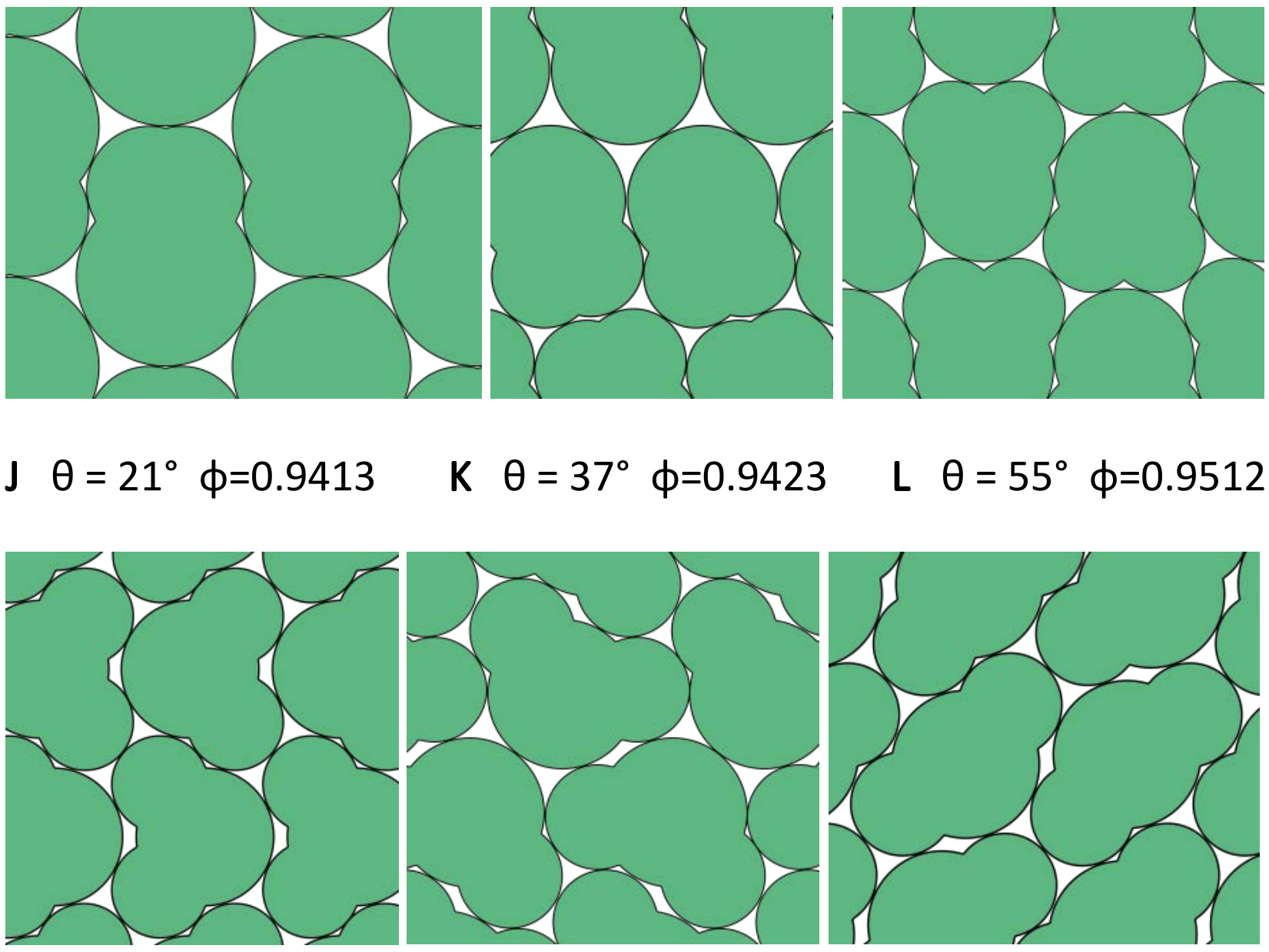

M $\theta$ = 99° $\phi$=0.9411 N $\theta$ = 119° $\phi$=0.9404 P $\theta$ = 158° $\phi$=0.9404

**Figure 8**. The structures corresponding to the packing fraction peaks as labelled in Fig. 7 for the trimer (R=0.7) for which the MCN is 3 for all values of $\theta$ except within the range $111^o < \theta < 143^o$ where the MCN = 4. Exactly at the angles corresponding to the crossover between MCN = 3 and 4 we observe an MCN = 5.

Having introduced the MCN as a means of characterizing concavity, we can use the trimer data to explore in detail two aspects of the MCN, namely i) does a change in MCN correlate with any change in the densest packings? and ii) do these optimal packings always make use of the pairwise arrangement with a maximum number of contacts (i.e. an MCN pair)? As shown in Fig. 9, the MCN changes for the trimer as we vary the bond angle. The MCN =3 for most of the $\theta$ range but increases to a value of 4 in the range $111^o < \theta < 143^o$. At the two special angles, $111^o$ and $143^o$, the MCN jumps to 5. With regards our first question, we find no obvious correlation between the MCN of a shape and the resulting crystal structure or its packing fraction of across the range of bond angles. We do find, however, signs of a systematic increase in the contact number (averaged over the relevant range of bond angles) as the MCN increases. To address our second question, we have examined each of the densest p2 structures and identified which make use of an MCN pair and which do not. These are indicated in Fig. 9 with those structures that do not use an MCN pair shaded in grey. We find that all structures associated with peaks in the packing fraction (and local maxima in the contact number) do employ an MCN pair. For the remainder of the trimers, the densest packings do not always employ an MCN pair. We conclude that for some shapes, the use of an MCN pair is disfavoured because it results in an overall decrease in the total number of contacts per particle.

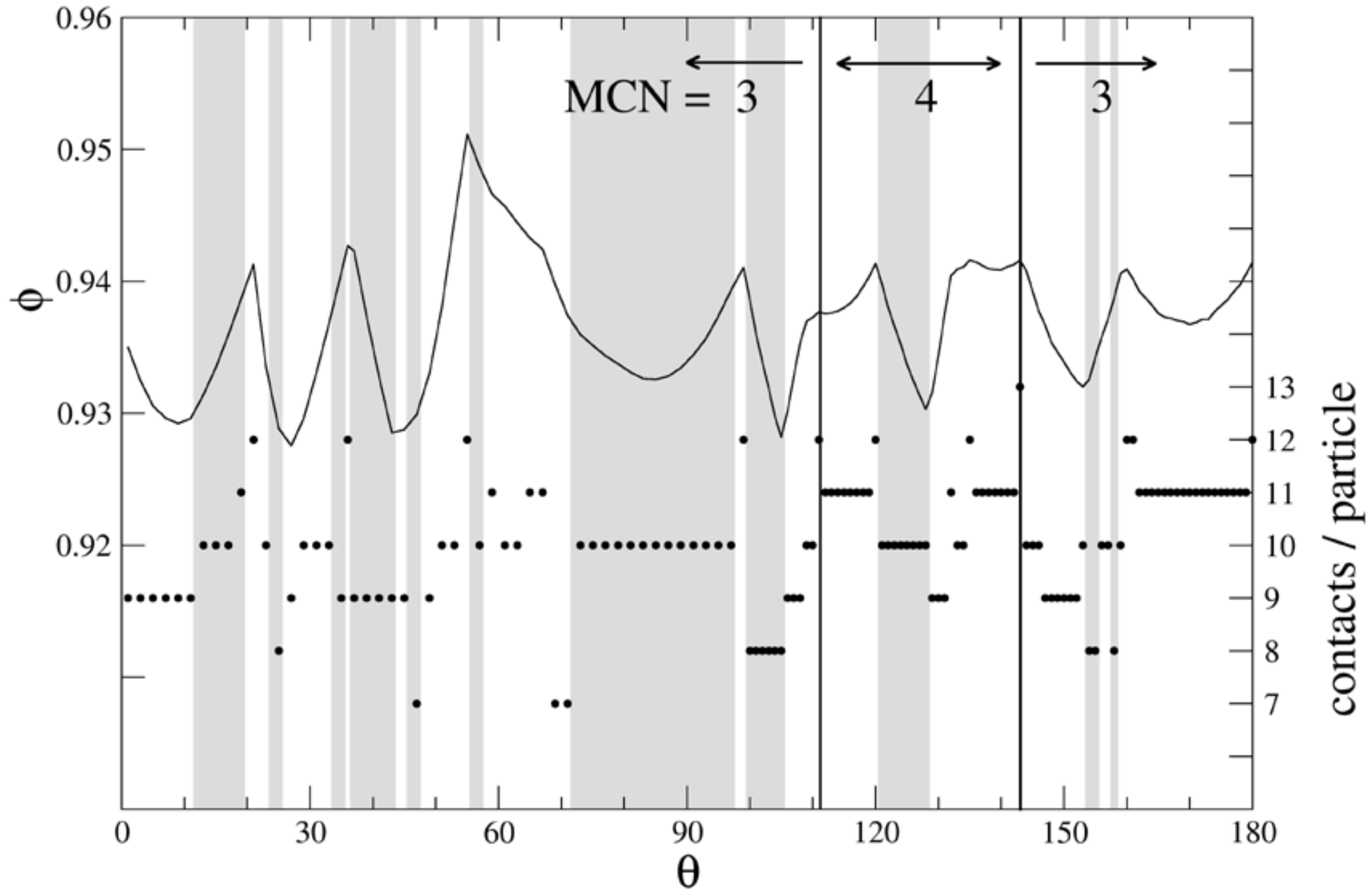

**Figure 9.** The maximum packing density of the p2 crystals of the R= 0.7 trimers and the number of contacts Z per particle, as indicated on the right hand axes, as a function of the bond angle. The shaded regions indicate values of θ for which the p2 crystal does not include pairs with MCN contacts. The value of the MCN varies with the bond angle as indicated by the arrows at the top of the figure.

In conclusion, we have found that the densest crystal packings for a range of dimers and trimers are dominated by just two space groups, p2 and p2gg, which produce structures essentially degenerate in terms of their packing fraction over much of the range of shapes studied. The optimization of the packing density appears to coincide with the optimization of the contact number per particle. Peaks in the maximum packing fraction with respect to shape change are strongly linked to singular jumps in the contact number and are always associated with the presence of MCN pairs in the packing. In general, however, the MCN pair does not always feature in the optimal packing. Finally, the concavity of the particle (as characterized by the MCN) is correlated with the average contact number in the sense that an increase in MCN increases the likelihood of producing a dense packing with an increase in contact number. Apart from this, concavities influence on crystal structures appears to be a rather general one, expressed in the limited number of space groups responsible for the densest packings.

## 4. Dense Amorphous Packings and Constraint Counting

### 3.1. Constraint Counting for Amorphous Packings of Anisotropic Particles

We now turn to the molecular packing in amorphous structures in order to see whether, unfettered from the symmetry constraints that crystal structures impose, particle shape plays a different role. Given the correlation already noted between concavity and contact number in the crystals, we shall begin by introducing the idea of constraint counting as a means of analysing their mechanical stability. In granular materials (which, for our purposes, are simply collections of congruent frictionless repulsive particles at zero temperature) a disordered jammed configuration is defined as an amorphous arrangement of particles one in which all degrees of freedom, save a small set of localised 'rattlers', are constrained [33]. It has been demonstrated [34] that such disordered jammed configurations are *isostatic,* meaning that the total number of particle contacts equals the total number of degrees of freedom, i.e.

$$Z_{iso}=2d_f \qquad (1)$$

where $Z_{iso}$ is the average number of contacts per particle in the isostatic configuration and $d_f$ is the number of degrees of freedom per particle (in 2D, $d_f$ = 2 or 3 for disks or anistropic particles, respectively). While the arguments for Eq.1 only rigorously apply to spherical particles, there is a general expectation that Eq. 1 holds for anisotropic particles as well, an expectation that has been called the 'isocounting conjecture' [35].

The validity of the extension of Eq. 1 to anisotropic particles has been found to depend on the details of the particle shape. According to this conjecture, $Z_{iso}$ = 6 for an anisotropic particle in 2D (since $d_f$ = 3). Schrek et al [36] found that the proposition that $Z_{iso}=2d_f$ is valid for the jamming point for dimers constructed from fusing two identical disks for all values of the aspect ratio but failed for ellipses for which Z was found to increase continuously from the value of 4 for the disks to only approach 6 asymptotically with increasing aspect ratio. The mechanical rigidity of the jammed ellipses, in spite of the apparent under-constraint, was explained by Donev et al [35] as a consequence of contributions to rigidity due to terms second order and higher in the orientational fluctuations. The striking difference between dimers and ellipses with the same aspect ratio suggests that whether a particle was concave (as in the case of the dimer) or convex , like the ellipse, has significant consequences with regards the nature of mechanical stability in the amorphous state.

### 3.2. Contact Statistics of Amorphous Packings

To generate amorphous packings of ansotropic shapes, we have made use of a standard molecular dynamics (MD) algorithm, LAMMPS [37], and particle models consisting of continuous potentials. The dimer and trimer shapes are still constructed as described in Fig. 1 except instead of treating the component centres as the centres of hard disks of fixed radius we now treat them as the centres of an isotropic Lennard-Jones interaction, where the pairwise interaction potential is given by

$$u(r) = 4\varepsilon\left[\left(\frac{\sigma_{ij}}{r}\right)^{12} - \left(\frac{\sigma_{ij}}{r}\right)^{6}\right] \qquad (2)$$

where $\varepsilon = 1$ and $\sigma_{11}=2.0$, $\sigma_{22} = 2R$ and $\sigma_{12} = 1.0+R$. The amorphous packings are generated by conjugate gradient minimization of the potential energy of an initial disordered configuration of particles generated by an extended MD run at a high temperature. The minimized amorphous configurations are referred to as ‘inherent structures’ (IS). In Fig. 11 we provide an example of an IS for the trimer (R=0.7, $\theta = 120^o$).

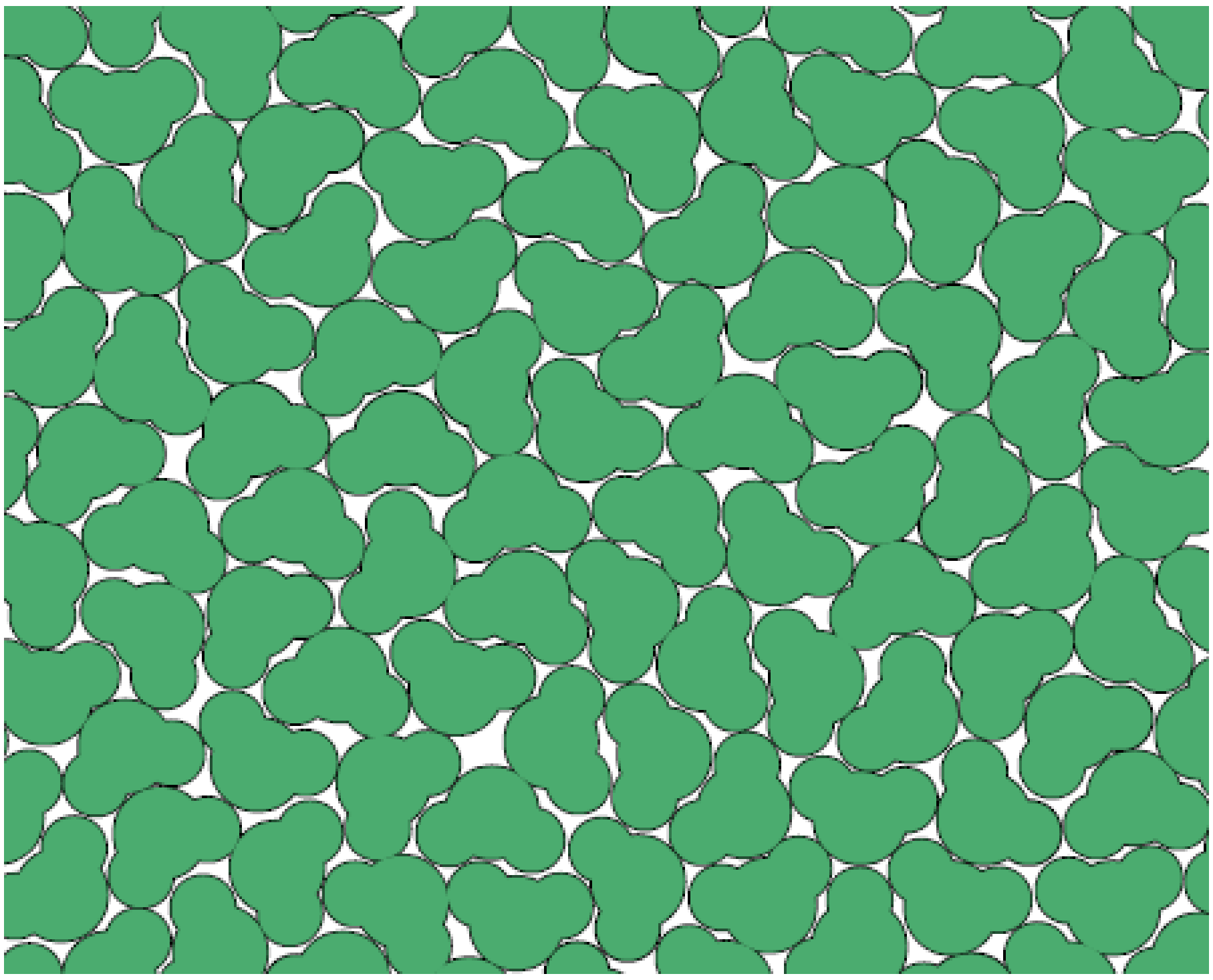

**Figure 10.** An example of an inherent structure of the trimer (R=0.7, $\theta = 120^o$).

To collect statistics concerning the contacts between particles we must define a contact for these particles. A contact is said to exist if two component centres of type i and j on different particles are separated by a distance $r \le 2^{1/6}\sigma_{ij}$. The difference between this definition of contact and that used for the hard particles in the previous Section complicates any direct comparison. Where we do contrast the amorphous and crystal contact statistics in the following discussion, we shall re-evaluate the crystal contact data using the analogous ‘softer’ contact condition as used for the Lennard-Jones interactions, i.e. a contact is registered in the crystal of the separation between centres $r \le 2^{1/6}(R_i + R_j)$ .

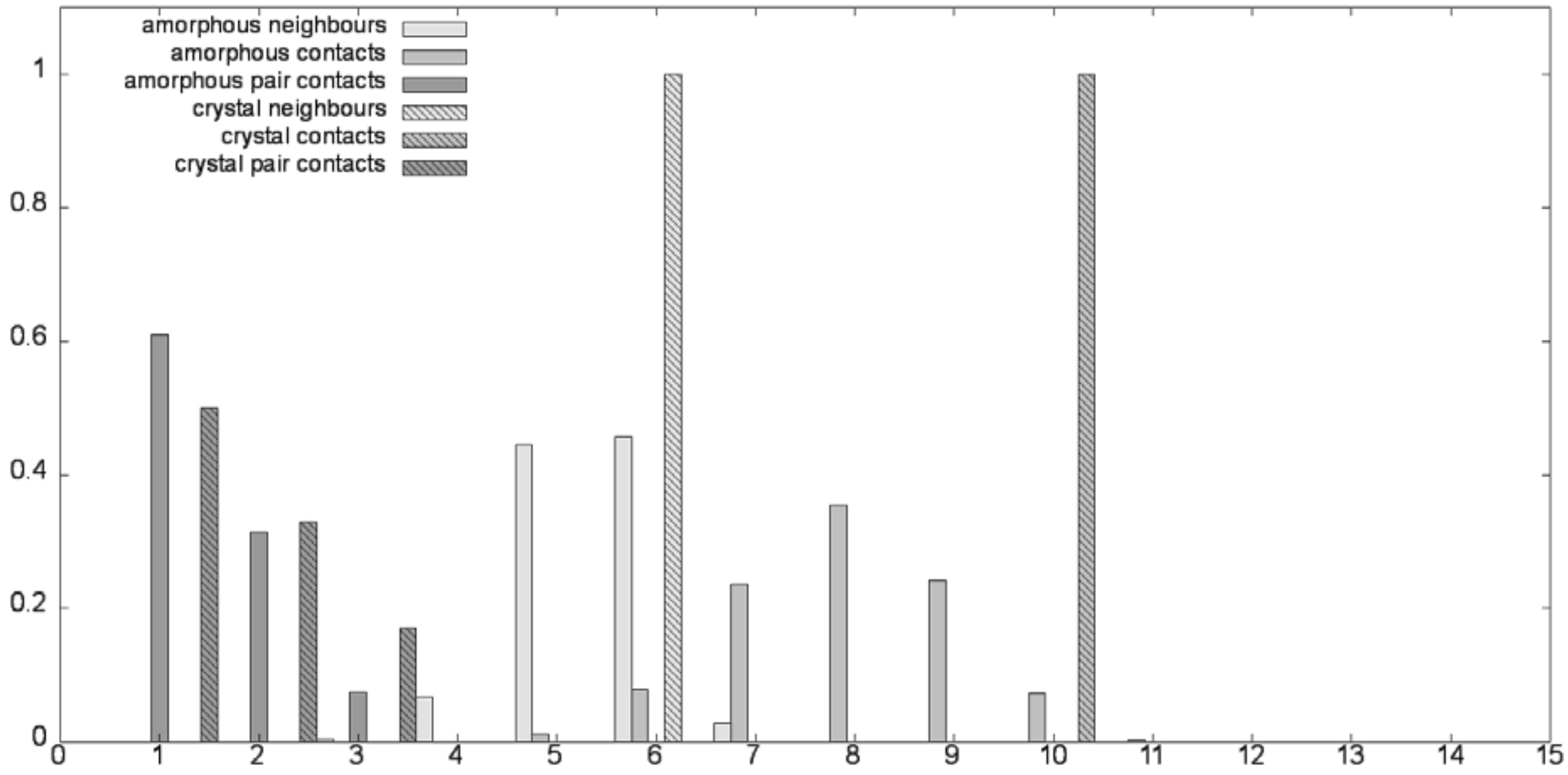

**Figure 11**. Distributions of the number of contacts per pair, the contacts per particle and the number of neighbours for amorphous packings of a dimer (R=0.7). The amorphous states were generated from a conjugant gradient minimizing of the energy from T=5.0 liquids.

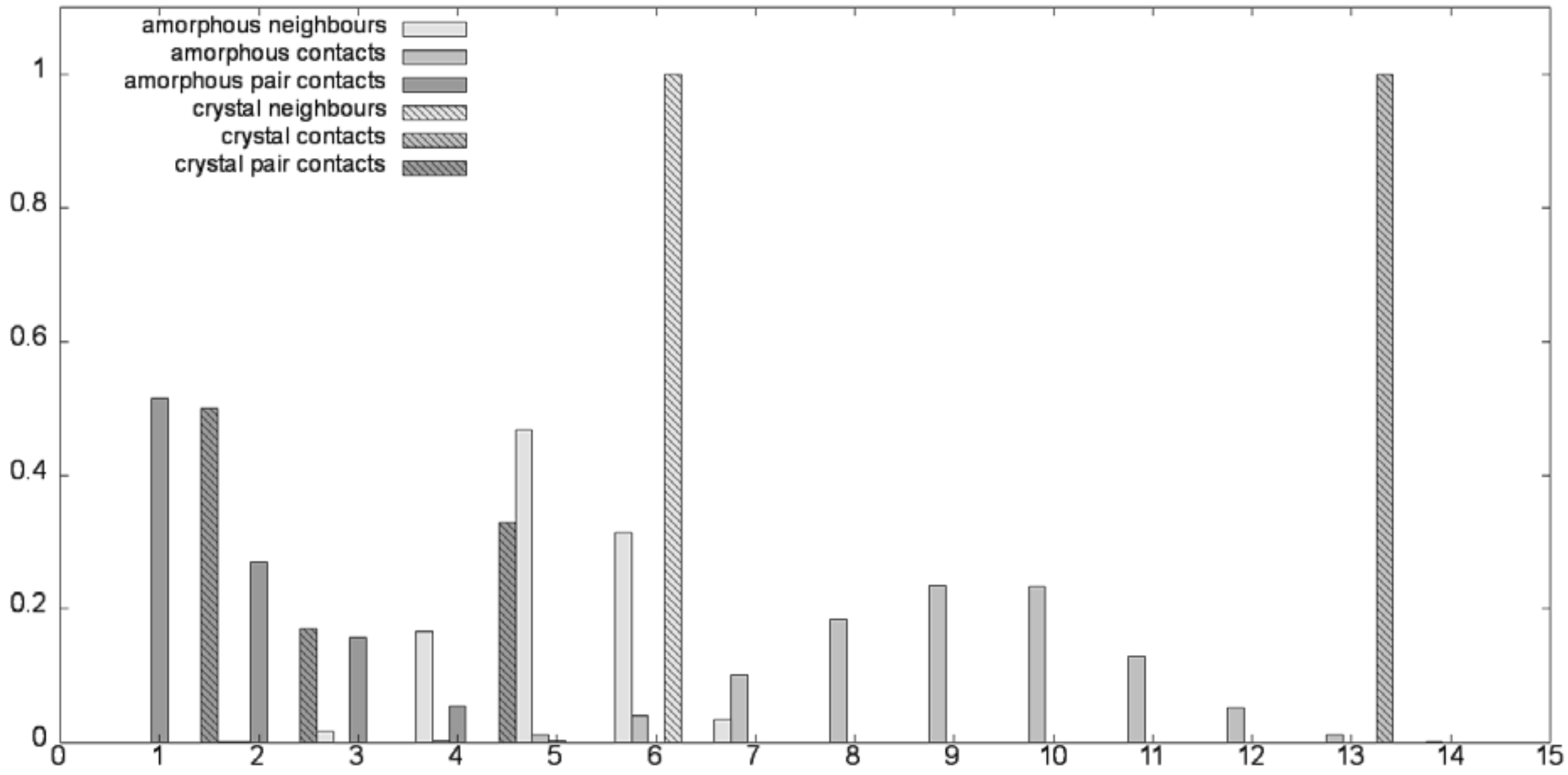

**Figure 12.** Distributions of the number of contacts per pair, the contacts per particle and the number of neighbours for amorphous packings of a trimer (R=0.7,θ=120°). Note that the crystal contact data here used the same definition of contact as used for the amorphous state, i.e. $r \le 2^{1/6}(R_i + R_j)$ . The amorphous states were generated from a conjugant gradient minimizing of the energy from T=5.0 liquids.

We have examined amorphous configurations of dimers and trimers for small disk radii R ranging from 0.6 to 0.9 and observe no significant variation in the statistics of contacts per pair, contacts per particle or coordination number. Contact statistics for a dimer (R=0.7) and a trimer (R=0.7, $\theta=120^o$) are presented in Figs. 11 and 12. We make a number of observations of the statistics presented in these figures.

i) The disorder of these configurations is evident in the distribution of the observed number of neighbours around the ideal value of 6 with the amorphous character of the dimer (Fig. 11) and the $\theta = 120^o$ trimer (Fig. 12) characterised by a distribution of nearest neighbour numbers.

ii) The number of contacts per particle is represented by a broad distribution whose upper bound is set by the contact number the crystal. We see a general increase in the mean contact number and the width of the distribution with the increasing number of sites used to define the particle so that over-constraint in some sense seems unavoidable. The broad distribution of contact numbers raises a number of interesting questions about the possible connection with heterogeneities in dynamical and mechanical response which we leave for the future.

iii) In contrast to the difference in the total contact numbers per particle, we find that the distribution of contact numbers *per pair* are similar for the crystal and amorphous structures in the dimer (Fig. 11) and the $\theta = 120^o$ trimer (Fig. 12). This suggests that the distribution in contact numbers results from a distribution in the local 'allocation' of different types of pairwise contacts rather than a global difference in the types of pairings.

As shown in Fig. 9, the $120^o$ trimer has an MCN = 4 and is an 'optimal' shape in the sense that it corresponds to a peak in the packing fraction with a crystal structure characterised by a large contact number (i.e. 12) and the presence of an MCN pair. How will the comparison between crystal and amorphous structures differ if we where to consider an 'ordinary' trimer? To address this question we have consider the $80^o$ trimer, a shape not associated with any packing fraction peak and one whose densest crystal packing does not use an MCN pair (see Fig. 9). In Fig. 13 we plot the distribution of pairwise contacts, coordination number and contact numbers for the amorphous solid and densest crystal for the $80^o$ trimer.  We find that the distribution of contact numbers in the amorphous solid is significantly reduced relative to that of the $120^o$ trimer and the crystal contact number no longer represents an upper bound on the amorphous distribution. Likewise, the distribution of pairwise contacts in the amorphous solid differs significantly from that of the crystal, notably with the clear presence of MCN pairs absent from the crystal. These results suggest that the contact number of the crystal and amorphous solids are correlated, even while the amorphous state is able to make more consistent use of the MCN pair.

The amorphous solid, by virtue of its multiplicity of local environments, allows some further insight into the relationship between the presence of the MCN pair and the associated contact number and local order. Does the presence of a MCN pair (or, more generally, a high contact pair) result in the reduction of the total number of contacts for the participating particles?  In Fig. 14 we plot the average number of contacts per particle and the average number of neighbours per particle as a function of the local maximum number of contacts per pair for each individual particle and its neighbours.

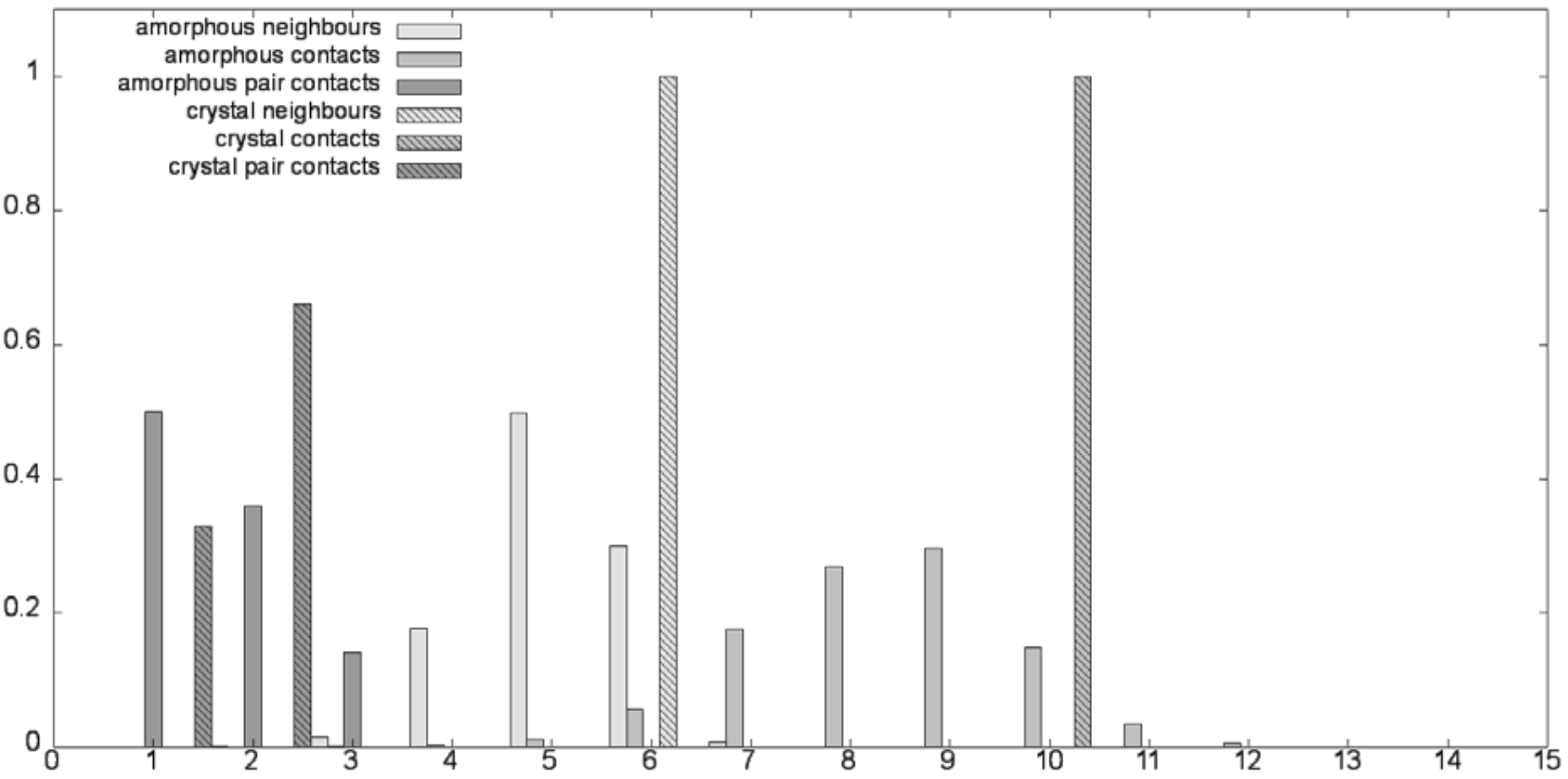

**Figure 13.** Distributions of the number of contacts per pair, the contacts per particle and the number of neighbours for amorphous packings of a trimer (R=0.7,θ=80°). Note that the crystal contact data here used the same definition of contact as used for the amorphous state, i.e. $r \le 2^{1/6}(R_i + R_j)$. The amorphous states were generated from a conjugant gradient minimizing of the energy from T=5.0 liquids. Note that the local MCN in the amorphous state clearly exceeds that of the crystal.

The data in Fig. 14 provide us with two unequivocal conclusions: i) there is no 'cost' for a high contact pair in the form of loss of overall number of local contacts; in fact, the number of contacts clearly increases with the maximum pairwise contact number, and ii) there is a clear trend for a decrease in the average coordination number as particles make use of the high contact pairings. The contrast in these two findings is interesting. The loss of coordination number suggests that high contact pairing tend to generate local disorder, a conclusion in agreement with their absence in crystal phases except for specific particle shapes. The increase in the contact number, on the other hand, suggests that there may be a local preference for these high contact pairings.

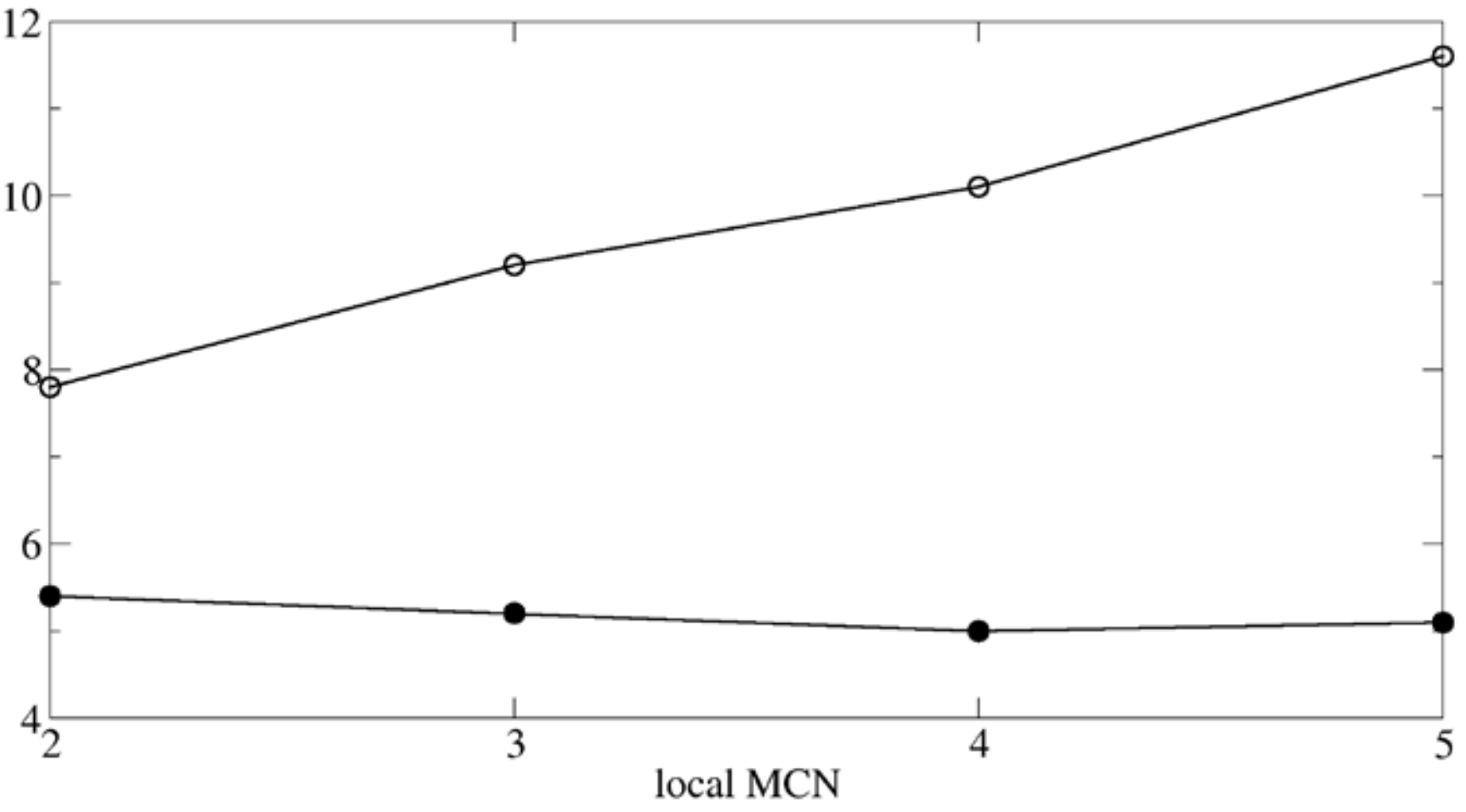

**Figure 14.** The average value of the number of contacts per particle (open) and number of nearest neighbours (filled) as a function of the largest single pairwise contact number (local MCN) exhibited by individual particles in the amorphous packing of the trimer (R=0.7, θ = 120°). Note that the contact number increases with increasing local MCN while the coordination number tends to decrease.

In the cases presented here, we are looking at systems where the constraints, in the case of the trimers, are clearly in excess of those associated with an isostatic state, i.e. the amorphous packing of concave particles is typically *hyperstatic*. The physical consequences of this over-constraint in amorphous systems of concave particles remains a largely open question. Here we demonstrate one straightforward feature of a hyperstatic state, namely that small clusters of particles can exhibit local rigidity without the need for global rigidity. Consider a hexagonal cluster of a triangular lattice made up of L layers (and a central particle) as depicted in Fig. 15. The number of particles in such a cluster is N = 1+3L(L+1), the number of degrees of freedom $d_f$=3N-3 (i.e. two translational and one rotational degree of freedom) and the number of neighbour pairs = $9L^2 + 3L$. Let Z be the average number of contacts per particle for the bulk. (Note that it is through this variable that the shape of the particle enters this treatment since, for a convex particle, Z is constrained to equal 6. The greater the difference Z-6, the greater the concavity of the particle.) It follows that the number of constraints $n_c$ in the cluster of L layers is $n_c = (Z/6)(9L^2+3L)$. The number of floppy modes is given by the difference

$$d_f - n_c = 9L^2(1-Z/6)+3L(3-Z/6)$$

which ≤ 0 (i.e. is rigid) for L = 1 (i.e. the local coordination shell) when Z ≥ 9. Since the dimers and trimers studied almost always satisfy this condition for Z (see Figs. 11-13) we predict that local coordination structures are will be rigid. This is different for the convex particles where rigidity arises only in the limit of large clusters. Local rigidity could be manifest in particularly slow relaxation dynamics. Ref. 38 found little difference in the density dependence of structural relaxation between dimers and ellipses with similar aspect ratio (although the fragility of both was found to increase monotonically with aspect ratio). In neither case, however, did the configurations exhibit contacts in excess of the isostatic prediction. Following the argument of Donev et al [35], both systems were actually isostatic – a conclusion consistent with the similarity in their relaxation kinetics.

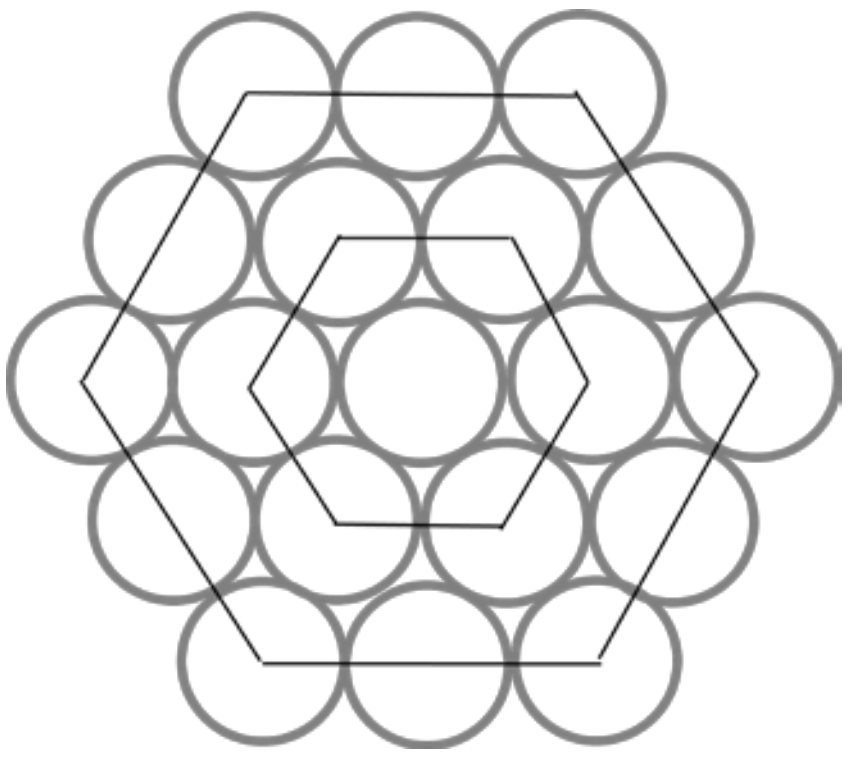

**Figure 15.** An hexagonal cluster of particles comprised of two hexagonal layers (indicated by lines), i.e. L =2.

Applying the same analysis to convex anisotropic particles we find that the number of floppy modes is 6L, i.e. a number of rotations proportional to the cluster surface area that are never constrained. It is the particle concavities and the additional constraints they introduce that, at least according to this simple analysis, pins these surface rotations.

## 5. Discussion

Our goal in this paper was to establish how the shape of concave particle shapes influence their packing. Our conclusions come with the qualification that they apply to the families of shapes generated by overlapping disks. It will be the task of future work to understand whether the discussion here applies to a broader family of concave shapes. When this packing in question is that of the densest crystal, we find that these crystals belong almost entirely to just two space groups: p2 and p2gg. The structures in these two space groups are frequently degenerate in density, despite significant structural differences.

| **2D Structure Data Base** [39] (expt) | **Data Base from ref. [25]** (theory) | **From this work** (theory) |
|---|---|---|
| p2 (54%) * | p2 (54%) * | p2* + p2gg (~ 96%) |
| p1 (17%) * | p2gg (20%) | |
| p2gg (9.7%) | pmg (10%) | |
| cm (5.3%) | p4 (6.5%) * | |
| pg (4.2%) | p3 (3.1%) * | |
| p6 (2.2%) | | |

**Table 1.** The frequency of various wallpaper groups in 2D crystals from this work, from a previous simulation study [25] for concave shapes generated by spherical harmonic perturbation of disks and from experimental data [39] on molecules adsorbed on crystalline surfaces. * indicates wallpaper groups that can accommodate chiral shapes.

In Table 1 we compare the frequency of wallpaper groups found in crystals structures from this study, a previous simulation study [25] in which concave particles were generated by circular harmonic perturbations of hard disks and from a compilation of experimental data for 2D molecular crystals adsorbed on solid surfaces [39]. There is a clear difference in the frequency of wallpaper groups. Each set of data explores a different space of shapes. The experimental data, for example, includes a large number of examples of extended alkane chains. The shapes from ref. [17] were generated stochastically from a space parametrized by the amplitude and wavelength of harmonic perturbations of a hard disk. The shapes from our study were based on the overlap of circular disks and so presumably some of the specific peculiarities of disk packings. What is striking is that, despite these differences in the family of shapes studied, the p2 + p2gg space groups dominate in each case (63.7% of the experimental structures and 74% of the ref. [25] structures). The particular preponderance of p2 structures is consistent with the principal of rotated –pair packings proposed in ref. [23]. The p2 group is one of the 4 space groups identified as ‘allowed’ for dense packing by the packing argument of Kitaigorodsky [30], i.e. that close packing requires a 6-fold coordination. What remains a puzzle is why the p2 group, specifically, so dominates the optimal packing. There appears to be a general packing benefit associated with the organization of a rotated pair on a simple lattice that we still don’t fully understand.

As we vary the particle shape, we find that the packing fraction is characterised by a sequence of cusp-like peaks, each peak associated with a particular ideal match between a shape and a particular packing arrangement. The peaks in packing fraction are associated with peaks in the number of contacts per particle supporting the general that the densest crystal corresponds to the maximum average number of contacts per particle. In this paper we introduced the maximum contact number (MCN) between a pair of congruent particles as a useful parameter for classifying concavity. We demonstrated that, in the crystal, the influence of the MCN was modest, limited to the observation that an increase in MCN was associated with an increasing likelihood of observing a high contact number. In the amorphous solid, however, the MCN may play a more significant role, with marked differences in the distribution of contact numbers observed between the 120$^o$ trimer (MCN=4) and the 80$^o$ trimer (MCN=3). Our results suggest that high contact pairing is likely to play a more general role in the structure and dynamics of amorphous phases than in the case of crystal phases. This conclusion represents a qualification of the general assumption of the importance of optimal pairing of molecules that dates back at least to the 1940 paper by Pauling and Delbrueck [40]. In the context of discussion of the stabilization of molecular association in biological systems, these authors argued that : ".. in order to achieve the maximum stability, the two molecules must have complimentary surfaces, lie die and coin..". Our results here make clear that this 'die and coin' type packing is often not used in optimizing the density in crystals while playing a generally more significant role in amorphous states.

We have demonstrated that the amorphous states of concave molecules are typically hyperstatic. This raises a range of open questions concerning the physical consequences of such over-constraint of particle motions including the effect of concavity on normal modes, the importance of local fluctuations for relaxation given the locality of rigidity, and the manner by which a liquid of concave particles approaches mechanical arrest on cooling. It is our hope that the study presented here provides a useful basis for this future research.

**Acknowledgements**

The authors would like to take acknowledge the important contributions of Jean-Pierre Hansen to the theory of simple liquids, the statistical mechanics of freezing and the study of supercooled liquids, contributions for which the authors are indebted. The authors also acknowledge the support of the Australian Research Council through the Discovery Project program.

**Appendix**

As the designation of 2D crystal structures in terms of their space groups may not be familiar to all readers, we have included here examples of those space groups mentioned in this paper. Each of the structures depicted corresponds to the densest packing of the trimer (θ=120$^o$ and R = 0.7) for that space group (and a particular choice of Wykoff site).

In Fig. 16 we provide an example of the simplest space group, p1, where a single particle is placed in each unit cell in any arbitrary orientation. In contrast, a pm structure has a plane of mirror symmetry, so either the particle lies on that mirror plane (Wyckoff site "a") with its own mirror plane aligned, or there are two particles per cell (Wyckoff site "c") related by the mirror.

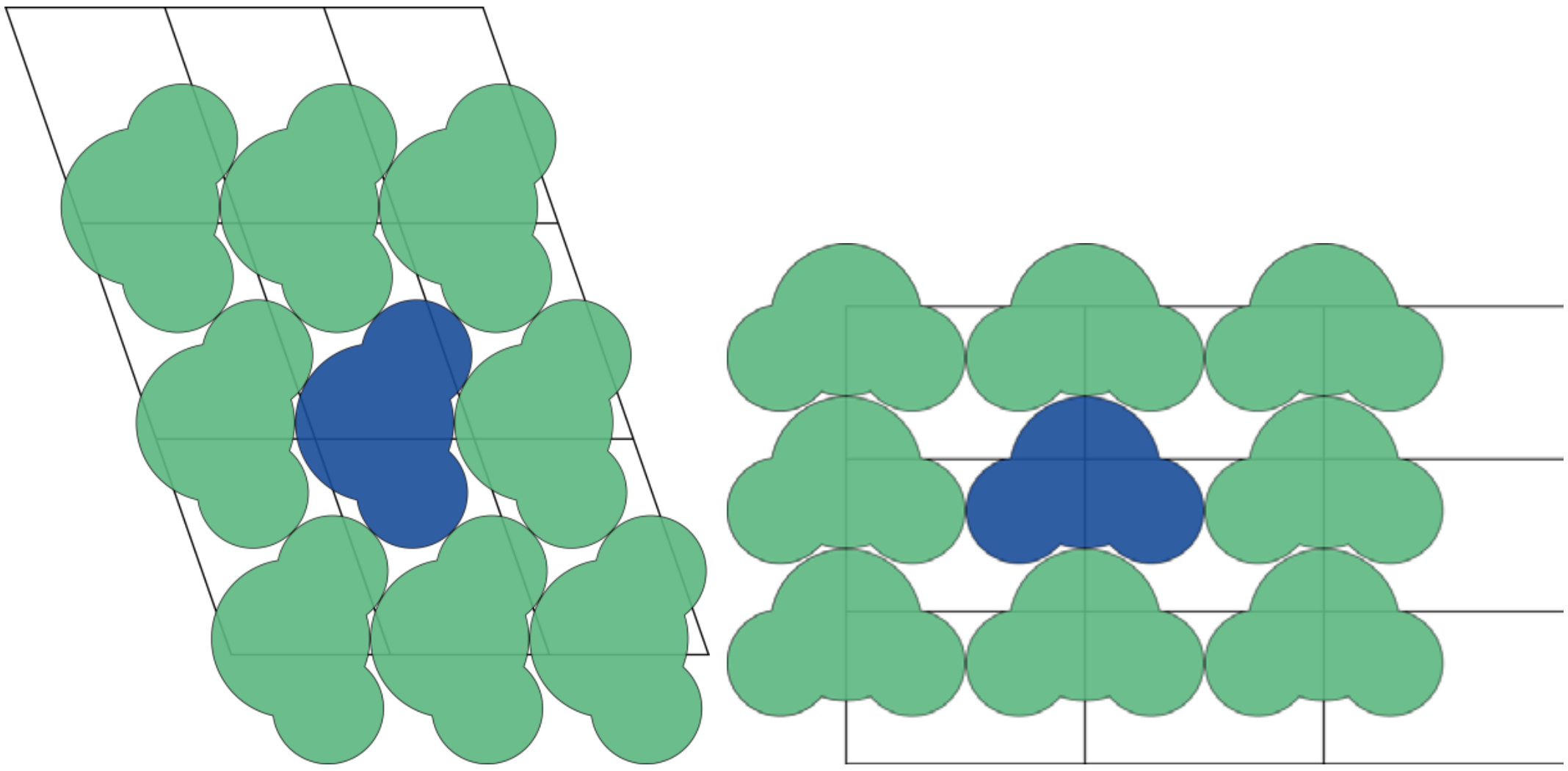

**Figure 16**. Examples of structures belonging to the p1(left) and pm (right) space groups. The blue particle(s) in each case represent the particle(s) within a single unit cell. Note that the pm space group has a plane of mirror symmetry, and the mirror plane of the achiral particles is aligned with it.

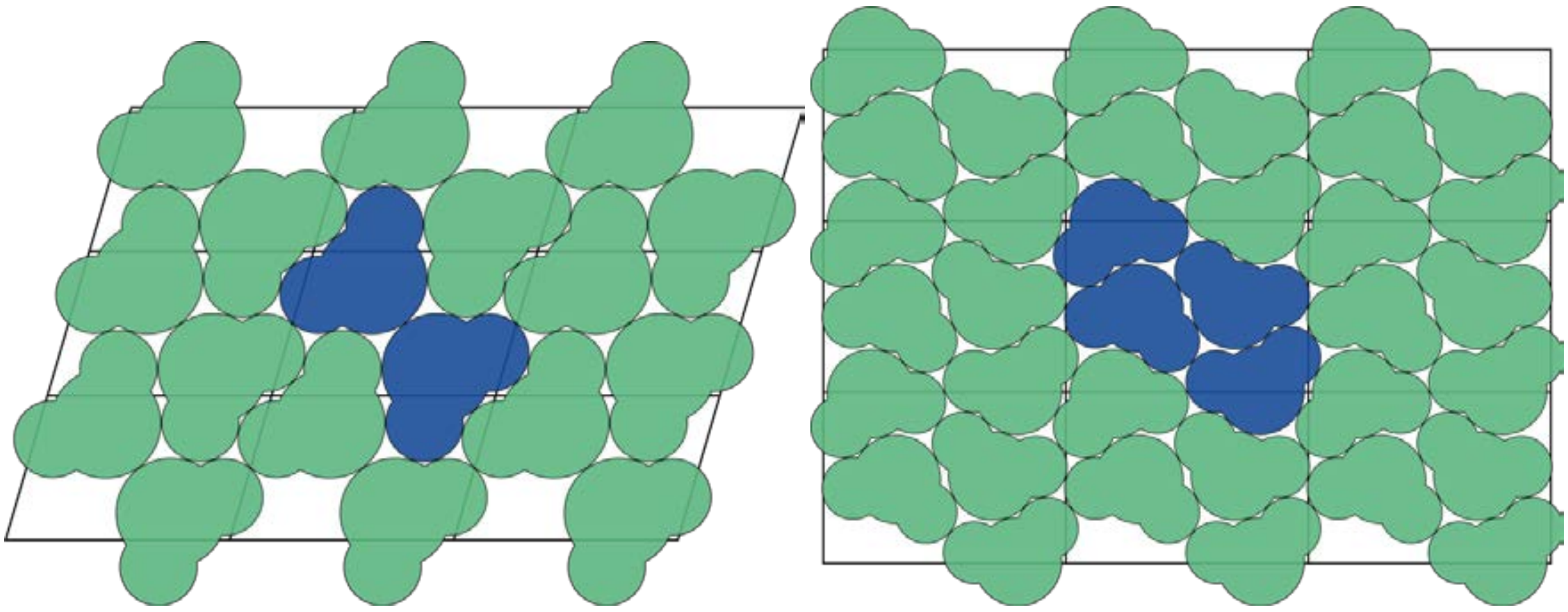

**Figure 17.** Examples of structures belonging to the p2 (left) and p2gg (right) space groups. The blue particles in each case represent the particles within a single unit cell.

Structural complexity is increased by increasing the number of particles in a unit cell and constraints imposed by the symmetry relations (see Fig. 17). The p2 structure consists of a pair of particles related by a 180° rotation. The p2gg structure has two pairs related by a 180° rotation, where the pairs are themselves related by glide reflections in two perpendicular directions.

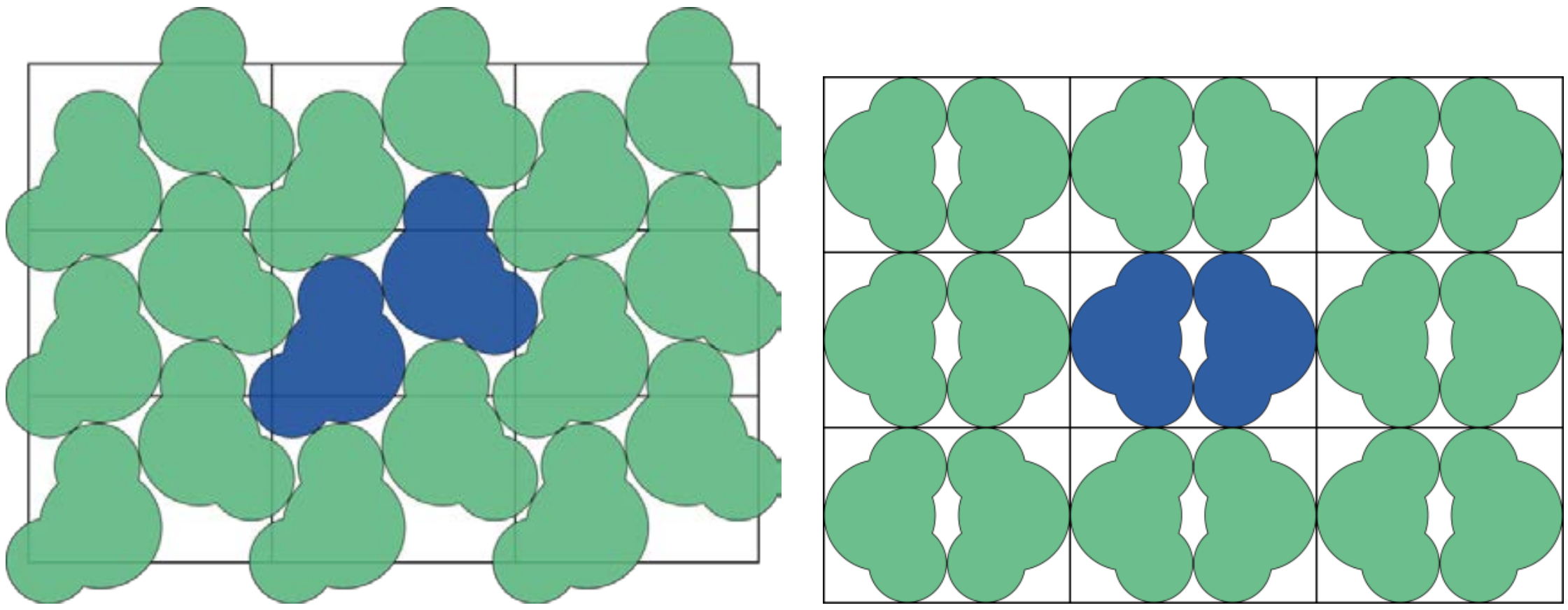

**Figure 18.** Examples of structures belonging to the pg (left) and p2mm (right) space groups.

In Fig. 18, the pg structure contains a glide reflection, so although the cell has a pair of particles, they are not necessarily in the "rotated pair" configuration that often occurs in densest packings. The p2mm has 180° rotational centres and therefore rotated pairs, but it also has two perpendicular reflections which impose additional constraints so that these particles only have four neighbours, and a low packing density.

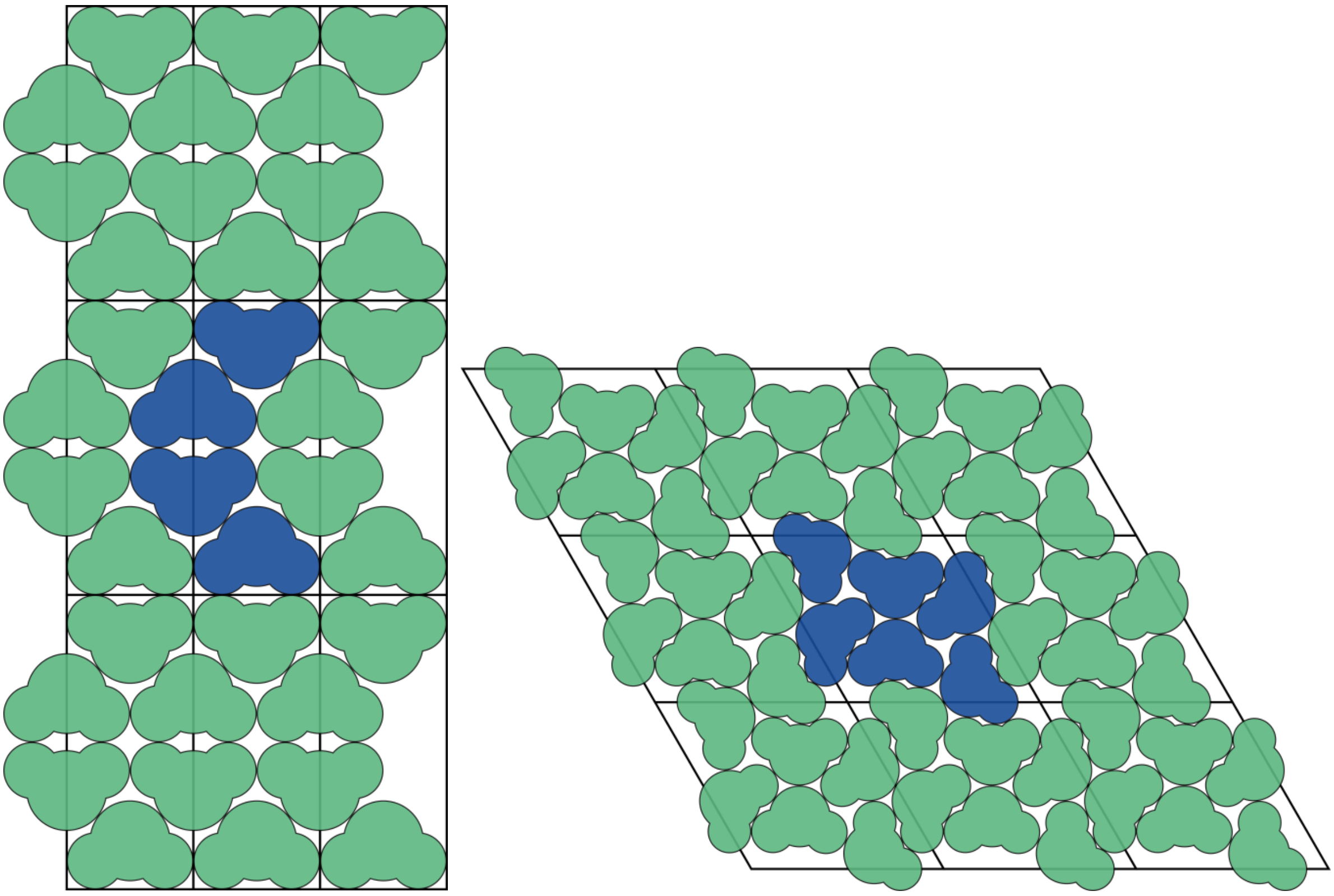

**Figure 19.** Examples of structures belonging to the c2mm (left) and p6 (right) space groups.

Most of the possible space groups never manage produce the densest packing. The examples of c2mm and p6 structures in Fig. 19 make clear how the extra symmetries of these structures can force voids into the crystal structure, even when the exact particle position and cell size is optimized within the symmetry.

## References

1. J.-L. Barat and J.-P. Hansen, *Basic Concepts for Simple and Complex Liquids* (Cambridge University Press, Cambridge, 2003).

2. F. J. Dyson and A. Lenard, J. Math. Phys. **8**, 423 (1967).

3. L. Verlet, Phys. Rev. **165**, 201 (1968).

4. T. Aste and D. Weaire, *The Pursuit of Perfect Packing* (Oxford University Press, London, 1966); T.C. Hales, Annals Math. **162**, 1065 (2005).

5. T. S. Hudson and P. Harrowell, J. Phys. Chem. B **112**, 8139 (2008); J. K. Kummerfeld, T. S. Hudson and P. Harrowell, J. Phys. Chem. B **112**, 10773 (2008); P. I. O'Toole and T. S. Hudson, J. Phys. Chem. C **115**, 19037 (2011).

6. A. B. Hopkins, F. H. Stillinger and S. Torquato, Phys. Rev. E **85**, 021130 (2012).

7. A. Donev, F. H. Stillinger, P. M. Chaikin and S. Torquato, Phys. Rev. Lett. **92**, 255506 (2004).

8. C. Ferreiro-Cordova and J. S. van Duijneveldt, J. Chem. Eng. Data **59**, 3055 (2014).

9. J. A. C. Veerman and D. Frenkel, Phys. Rev. A **45**, 5632 (1992).

10. S. Torquato and Y. Jiao, Phys. Rev. E **80**, 041104 (2009); E. R. Chen, D. Klotsa, M. Engel, P. F. Damasceno and S. C. Glotzer, Phys. Rev. X **4**, 011024 (2014).

11. J. de Graaf, R. van Roij and M. Dijkstra, Phys. Rev. Lett. **107**, 155501 (2011).

12. M. Dennison, K. Milinkovic and M. Dijkstra, J. Chem. Phys. **137**, 044507 (2012).

13. K. Milinkovic, M. Dennison and M. Dijkstra, Phys. Rev. E **87**, 032128 (2013).

14. P. F. Damasceno, M.Engel and S. C. Glotzer, ACS Nano **6**, 609 (2012).

15. V. N. Manoharan, M. T. Elsesser and D. J. Pine, Science **301**, 483 (2003).

16. P. M. Johnson, C. M. van Kats and A. van Blaaderen, Langmuir **21**, 11510 (2005).

17. D. Zerrouki, J. Baudry, D. Pine, P. Chaikin and J. Bibette, Nature **455**, 380 (2008).

18. D. J. Kraft, J. Groenewold and W. K. Kegel, Soft Matt. **5**, 3823 (2009).

19. J. R. Wolters, G. Avvisati, F. Hagemans, T. Vissers, D. J. Kraft, M. Dijkstra and W. K. Kegel, Soft Matt. **11**, 1067 (2015)

20. R. Rice, R. Roth and C. P. Royall, Soft Matt. **8**, 1163 (2012).

21. S. Scanna, W. T. M. Irvine, P. M. Chakin and D. J. Pine, Nature **464**, 575 (2010).

22. D. J. Ashton, S. J. Ivell, R. P. A. Dullens, R. L. Jack, N. B. Wilding and D. G. A. L. Aarts, arXiv:1412.1596v1 (2014).

23. S. Torquato and Y. Jiao, Phys. Rev. E **86**, 011102 (2012).

24. CEGO, et al Europhys. Lett. **98**, 44008 (2012).

25. D. F. Tracey, Honours Thesis, University of Sydney (2013); D. F. Tracey, A. Widmer-Cooper and T. S. Hudson, in preparation.

26. T. S. Hudson and P. Harrowell, J. Phys.: Cond. Matt. **23**, 194103 (2011).

27. N. T. Elias and T. S. Hudson, J. Phys.: Conf. Series **402**, 012005 (2012).

28. H. Wondratschek, *International Tables of Crystallography Vol. A* (ed. Th. Hahn) (International Union of Crystallography, 2006), pp. 732-740.

29. J. W. Steed, CrystEngComm **5**, 169 (2003).

30. A. I. Kitaigorodsky, *Molecular Crystals and Molecules* (Academic Press, New York, 1973)

31. C. P. Brock and J. D. Dunitz, Chem. Mater. **6**, 1118 (1994).

32. G. Kuperberg and W. Kuperberg, Discrete Comput. Geom. **5**, 389 (1990).

33. A. Baule and H. A. Makse, Soft Matt. **10**, 4423 (2014).

34. A. V. Tkachenko and T. A. Witten, Phys. Rev. E **60**, 687 (1999).

35. A. Donev, R. Connelly, F. H. Stillinger and S. Torquato, Phys. Rev. E **75**, 051304 (2007).

36. C. F. Schreck, N. Xu and C. S. O'Hern, Soft Matt. **6**, 2960 (2010).

37. S. Plimpton, J. Comp. Phys, **117**, 1 (1995).

38. T. Shen, C. Schreck, B. Chakraborty, D. E. Freed and C. S. O'Hern, Phys. Rev. E **86**, 041303 (2012).

39. K. E. Plass, A. L. Grzesiak and A. J. Matzger, Accounts Chem. Res. **40**, 287 (2007).

40. L. Pauling and M. Delbrueck, Science **92**, 77 (1940).